\documentclass[english,times]{iopart}
\usepackage{amssymb}
\usepackage{graphicx}
\usepackage{esint}

\usepackage{iopams}
\usepackage{setstack}

\begin{document}

\title{Thermodynamic cycle in a cavity optomechanical system}

\author{Hou Ian}

\address{Institute of Applied Physics and Materials Engineering, University
of Macau, Macau}
\begin{abstract}
A cavity optomechanical system is initiated by a radiation pressure
of a cavity field onto a mirror element acting as a quantum resonator.
This radiation pressure can control the thermodynamic character of
the mirror to some extent, such as cooling its effective temperature.
Here we show that by properly engineering the spectral density of
a thermal heat bath that interacts with a quantum system, the evolution
of the quantum system can be effectively turned on and off. Inside
a cavity optomechanical system, when the heat bath is realized by
a multi-mode oscillator modeling of the mirror, this on-off effect
translates to infusion or extraction of heat energy in and out of
the cavity field, facilitating a four-stroke thermodynamic cycle. 
\end{abstract}

\noindent{\it Keywords\/}: {cavity optomechanical systems, quantum control, quantum thermodynamics}

\maketitle

\section{Introduction}

The study of cavity electrodynamics roughly began with the discovery
of the Fabre-Perot inteferometer, in which two transflective side
mirrors sandwich an optical cavity of fixed length, thereby trapping
an optical cavity field of designate wavelength inside. When one of
the side mirrors is allowed to oscillate, usually by depositing a
transflective surface on a micro cantilever, the trapped cavity field
will interact with the movable mirror through radiation pressure~\cite{meystre85}
and other effects induced by the variable cavity length~\cite{kippenberg08}.

These kinds of controllable interactions provide some degrees of manipulation
to the movable mirror, thus opening the field of cavity optomechanics.
In particular, extensive studies have been conducted during the last
decade on how to cool down the effective temperature of the mirror~\cite{cooling1,cooling2,cooling3,cooling4,cooling5}.
Recently, studies on cavity optomechanical systems have found a wide
range of applications such as quadrature squeezing of polariton~\cite{ian08},
generation of Kerr nonlinearity~\cite{gong09}, and distant state
entanglement~\cite{wallquist10,joshi12}. However, the quantum thermodynamic
aspect of a cavity optomechanical system is less touched upon. In
this article, we study the thermodynamic evolution of the cavity field
under the influence of radiation pressure fedback from the mirror.

Generally speaking, when a heat bath modeled on an abstract manifold
is coupled to a quantum multi-level system, thermodynamic adiabatic
processes can be identified, during which work can flow in and out
of the heat bath~\cite{quan05}. Further, if the manifold is assumed
to be spin systems with particular temperature gradients, thermodynamic
machines can be facilitated~\cite{tonner05,youssef09}.

In the following sections, we show that, in a cavity optomechanical
system, these thermodynamic processes can be realized on the cavity
field. When the mirror is modeled as a multi-mode quantum oscillator,
it can play the role of heat bath that thermalizes the cavity system
according to the spectral density of states of the modeled oscillator.
In other words, when the spectral density is so specified, the ensemble
average energy of the cavity system evolves over time in the form
as a square wave, giving off an on-off effect about the interaction
between the system and the heat bath. The jumping of the energy up
and down on the square wave matches with the diabatic processes during
which heat is either infused into or extracted from the system. The
time during which the energy stays fixed designates the adiabatic
processes where no heat is transferred but work is done on the cavity
field.

To understand this complex process more clearly, we start our discussion
of quantum thermodynamics in Sec.~\ref{sec:Single-mode} below by
studying the simple interaction between a single-mode oscillator and
a two-level system. The ensemble average energy of the two-level system
is shown to be oscillating. The study is then expanded to the interaction
between a multi-mode oscillator and a two-level system in Sec.~\ref{sec:Multi-mode},
in which the on-off interaction effect is demonstrated. Thereafter,
modeling the heat bath on the multi-mode oscillator, the four processes
of the thermodynamic cycle are identified and the relevant spectral
density is given in Sec.~\ref{sec:Thermodynamic-Cycle}. The conclusion
is given in Sec.~\ref{sec:Conclusion}.

\section{Single-mode oscillating effect\label{sec:Single-mode}}

We consider a two-level system $\sigma_{z}=\left|e\right\rangle \left\langle e\right|-\left|g\right\rangle \left\langle g\right|$
as the main system and a single-mode oscillator $\{a,a^{\dagger}\}$
as the controller with a heat reservoir. Let their interaction be
the usual dipole-field coupling, then the total Hamiltonian is ($\hbar=1$)
\begin{equation}
H=\Omega\sigma_{z}+\omega a^{\dagger}a+\eta(a+a^{\dagger})\sigma_{z}.\label{eq:Ham_single}
\end{equation}
Further, let $\left|\psi_{n}^{e}\right\rangle $ ($\left|\psi_{n}^{g}\right\rangle $)
be the eigenstate of the controller associated with the excited state
$\left|e\right\rangle $ (ground state $\left|g\right\rangle $) of
the system, where $n$ designates the Fock number of the oscillator.
The tensor product $\left|e,\psi_{n}^{e}\right\rangle =\left|e\right\rangle \otimes\left|\psi_{n}^{e}\right\rangle $
describe the eigenstate of the combined system and controllor. Applying
the Hamiltonian (\ref{eq:Ham_single}) to this product state, we get
\begin{eqnarray}
H\left|e,\psi_{n}^{e}\right\rangle  & = & \left[\Omega+\omega a^{\dagger}a+\eta(a+a^{\dagger})\right]\left|e,\psi_{n}^{e}\right\rangle \nonumber \\
 & = & \left[\Omega+\omega\left(a^{\dagger}+\frac{\eta}{\omega}\right)\left(a+\frac{\eta}{\omega}\right)-\frac{\eta^{2}}{\omega}\right]\left|e,\psi_{n}^{e}\right\rangle .
\end{eqnarray}

That means, when the system stays in the excited state, the part of
Hamiltonian that determines the evolution of the controller is effectively
a displaced oscillator 
\begin{equation}
H^{e}=\omega A_{e}^{\dagger}A_{e}-\frac{\eta^{2}}{\omega}\label{eq:eff_Ham_e}
\end{equation}
where $A_{e}=a+\eta/\omega$. In other words, while interacting with
the excited system, the eigenstate of the controller is a displaced
Fock state: $\left|\psi_{n}^{e}\right\rangle =D(\mbox{\ensuremath{\frac{\eta}{\omega}}})\left|n\right\rangle $,
where $D(\mbox{\ensuremath{\frac{\eta}{\omega}}})=\exp\left\{ \mbox{\ensuremath{\frac{\eta}{\omega}}}(a^{\dagger}-a)\right\} $
denotes the displacement operator. Associated with this eigenstate,
the eigenvalue for the effective Hamiltonian~(\ref{eq:eff_Ham_e})
is then
\begin{equation}
H^{e}\left|\psi_{n}^{e}\right\rangle =\epsilon_{n}^{e}\left|\psi_{n}^{e}\right\rangle =\left(n\omega-\frac{\eta^{2}}{\omega}\right)\left|\psi_{n}^{e}\right\rangle 
\end{equation}
and the total eigenenergy of the combined system and controllor is
\begin{equation}
E_{n}^{e}=\Omega+n\omega-\frac{\eta^{2}}{\omega}.
\end{equation}

Following the same considerations, the system ground state $\left|g\right\rangle $
is associated with the inversely displaced Fock state $\left|\psi_{n}^{g}\right\rangle =D(-\frac{\eta}{\omega})\left|n\right\rangle $
of the controller. The effective Hamiltonian for the controller is
\begin{equation}
H^{g}=\omega A_{g}^{\dagger}A_{g}-\frac{\eta^{2}}{\omega}\label{eq:eff_Ham_g}
\end{equation}
with $A_{g}=a-\eta/\omega$, for which the total eigenenergy differs
from the excited state only by the sign of the system eigenenergy,
i.e.
\begin{equation}
E_{n}^{e}=-\Omega+n\omega-\frac{\eta^{2}}{\omega}.
\end{equation}

We can now consider the system dynamics for its coupling to the controller
thermal reservoir. Assume that the initial state has the controllor
retain a Fock number $n$ and the corresponding density matrix is
in a thermal equilibrium with Bernoulli distribution
\begin{equation}
\rho(0)=P_{e}\left|e,\psi_{n}^{e}(0)\right\rangle \left\langle e,\psi_{n}^{e}(0)\right|+P_{g}\left|g,\psi_{n}^{g}(0)\right\rangle \left\langle g,\psi_{n}^{g}(0)\right|.\label{eq:ini_density_mat}
\end{equation}
The Hamiltonian~(\ref{eq:Ham_single}) drives the evolution of the
system and the controller separately according to what we discussed
above, i.e. 
\begin{eqnarray}
\rho(t) & = & e^{-iHt}\rho(0)e^{iHt}\nonumber \\
 & = & P_{e}\left|e,\psi_{n}^{e}(t)\right\rangle \left\langle e,\psi_{n}^{e}(t)\right|+P_{g}\left|g,\psi_{n}^{g}(t)\right\rangle \left\langle g,\psi_{n}^{g}(t)\right|\label{eq:rho_sm}
\end{eqnarray}
where we have denoted $\left|\psi_{n}^{e}(t)\right\rangle =e^{-iH^{e}t}\left|\psi_{n}^{e}(0)\right\rangle $
and $\left|\psi_{n}^{g}(t)\right\rangle =e^{-iH^{g}t}\left|\psi_{n}^{g}(0)\right\rangle $.

The energies stored in the system $S$ and the controller $C$ varies
with time according to the initial Bernoulli distribution and the
system parameters but their total energy remains static if the density
matrix starts off from the initial state given in Eq.~(\ref{eq:ini_density_mat}).
Taking $\left\{ \left|e\right\rangle ,\left|g\right\rangle \right\} $
as the basis of the system and $\left\{ \left|\psi_{n}^{e}\right\rangle ,\left|\psi_{n}^{g}\right\rangle \right\} $
as the basis of the controllor, we can verify the constancy of the
total energy, i.e. 
\begin{eqnarray}
\left\langle H(t)\right\rangle  & = & \mbox{tr}_{S+C}\left(\rho(t)\left[\Omega\sigma_{z}+\omega a^{\dagger}a+\eta(a+a^{\dagger})\sigma_{z}\right]\right)\nonumber \\
 & = & \Omega(P_{e}-P_{g})+P_{e}\epsilon_{n}^{e}+P_{g}\epsilon_{n}^{g}.
\end{eqnarray}

However, since the controllor acting as a heat reservoir has constant
influx or outflux of thermal energy to and from the two-level system,
the energy of the system and its interaction with the controllor will
not remain constant. The average taken over the reduced density matrix
of the controllor gives
\begin{eqnarray}
\left\langle \Omega\sigma_{z}+\eta(a+a^{\dagger})\sigma_{z}\right\rangle _{C} & = & P_{e}\left\langle \psi_{n}^{e}\left|\Omega+\eta\left(a_{e}(t)+a_{e}^{\dagger}(t)\right)\right|\psi_{n}^{e}\right\rangle \sigma_{z}\nonumber \\
 &  & +P_{g}\left\langle \psi_{n}^{g}\left|\Omega+\eta\left(a_{g}(t)+a_{g}^{\dagger}(t)\right)\right|\psi_{n}^{g}\right\rangle \sigma_{z},\label{eq:avg_energy}
\end{eqnarray}
where $a_{e}(t)$ and $a_{g}(t)$ are the annhilation operators in
the Heisenberg picture 
\begin{eqnarray}
a_{e}(t) & = & e^{iH^{e}t}ae^{-iH^{e}t},\\
a_{g}(t) & = & e^{iH^{g}t}ae^{-iH^{g}t}.
\end{eqnarray}
of the controllor following the evolutions of the excited state and
ground state respectively. The Hamiltonians~(\ref{eq:eff_Ham_e})
and (\ref{eq:eff_Ham_g}) can be then considered as what are effectively
driving the dynamics of the controller since $\left[H(t),a_{e}(t)\right]=\left[H^{e}(t),a_{e}(t)\right]$
and $\left[H(t),a_{g}(t)\right]=\left[H^{g}(t),a_{g}(t)\right]$.

The time-dependent operators can be expressed explicitly by their
Heisenberg equations with respect to these Hamiltonians:
\begin{eqnarray}
\dot{a}_{e}(t) & = & -i\omega a_{e}(t)-i\eta,\label{eq:Hei_a_e}\\
\dot{a}_{g}(t) & = & -i\omega a_{g}(t)+i\eta,\label{eq:Hei_a_g}
\end{eqnarray}
for which the system-controller interaction $\eta$ is essentially
a driving of the level populations towards opposite directions for
the two system levels. We will see in the next section that this driving
translates to energy transfers in and out of the system levels. Substituting
the formal solutions to Eqs.~(\ref{eq:Hei_a_e})-(\ref{eq:Hei_a_g})
into (\ref{eq:avg_energy}), we find
\begin{eqnarray}
\left\langle \Omega\sigma_{z}+\eta(a+a^{\dagger})\sigma_{z}\right\rangle _{C}= &  & \Omega\sigma_{z}+\nonumber \\
\sum_{\gamma\in\{e,g\}}P_{\gamma}\eta\left\langle \psi_{n}^{\gamma}\right|a(0)e^{-i\omega t} &  & +a^{\dagger}(0)e^{i\omega t}+(-1)^{\gamma}\frac{2\eta}{\omega}(1-e^{-i\omega t})\left|\psi_{n}^{\gamma}\right\rangle 
\end{eqnarray}
where we let $\gamma$ be either 1 to indicate the positive sign for
the excited state or 0 to indicate the negative sign for the ground
state. The ensemble average of the operator $a^{\dagger}$ at the
initial state can be written as a c-number with amplitude $\alpha$
and phase $\phi$
\begin{equation}
\left\langle a^{\dagger}(0)\right\rangle =\left\langle \psi_{n}^{e}|a^{\dagger}(0)|\psi_{n}^{e}\right\rangle +\left\langle \psi_{n}^{g}|a^{\dagger}(0)|\psi_{n}^{g}\right\rangle =\alpha\cos\phi.
\end{equation}
Therefore, the average energy $\left\langle \Omega\sigma_{z}+\eta(a+a^{\dagger})\sigma_{z}\right\rangle _{C}$
becomes an oscillating value 
\begin{equation}
\left[\Omega+2\eta\alpha\cos(\omega t-\phi)+\frac{2\eta^{2}}{\omega}(P_{e}-P_{g})(\cos\omega t-1)\right]\sigma_{z},\label{eq:avg_single}
\end{equation}
where the direction of the oscillation depends on the system state.

\section{Multi-mode square wave on-off effect\label{sec:Multi-mode}}

We now extend the concepts introduced in the last section to the case
of a multi-mode oscillator. The Hamiltonian~(\ref{eq:Ham_single})
becomes
\begin{equation}
H=\Omega\sigma_{z}+\sum_{j}\omega_{j}a_{j}^{\dagger}a_{j}+\sum_{j}\eta_{j}(a_{j}+a_{j}^{\dagger})\sigma_{z},\label{eq:Ham_multi}
\end{equation}
where the system part $H_{S}=\Omega\sigma_{z}$ stays identical while
the controller part $H_{C}=\sum_{j}\omega_{j}a_{j}^{\dagger}a_{j}$
and the interaction part $H_{I}=\sum_{j}\eta_{j}(a_{j}+a_{j}^{\dagger})\sigma_{z}$
extends to the summation over all modes indexed by $j$. The associated
eigenstate for the excited system becomes a tensor product 
\begin{equation}
\left|e,\left\{ \psi_{n_{j}}^{e}\right\} \right\rangle =\left|e\right\rangle \prod_{\otimes j}\left|\psi_{n_{j}}^{e}\right\rangle 
\end{equation}
over the Fock states of all the photonic modes of the field.

Applying Eq.~(\ref{eq:Ham_multi}) to the eigenstate, we find the
effective Hamiltonian for the multi-mode controller $\left|\left\{ \psi_{n_{j}}^{e}\right\} \right\rangle $
to be
\begin{equation}
H^{e}=\sum_{j}\left(\omega_{j}A_{e,j}^{\dagger}A_{e,j}-\frac{\eta_{j}^{2}}{\omega_{j}}\right),
\end{equation}
where the displaced annhilation operator is now distinguished for
each mode
\begin{equation}
A_{e,j}=D^{-1}\left(\frac{\eta_{j}}{\omega_{j}}\right)a_{j}D\left(\frac{\eta_{j}}{\omega_{j}}\right).
\end{equation}
An identical procedure can be applied to the ground state, for which
the index $e$ in the equations above will be replaced by $g$.

When the two-level system retains its Bernoulli distribution, the
associated density matrix here only differs from Eq.~(\ref{eq:rho_sm})
by the expressions in the eigenstates, i.e. we can verify
\begin{eqnarray}
\rho(t) & = & P_{e}\left|e,\left\{ \psi_{n_{j}}^{e}(t)\right\} \right\rangle \left\langle e,\left\{ \psi_{n_{j}}^{e}(t)\right\} \right|\nonumber \\
 &  & +P_{g}\left|g,\left\{ \psi_{n_{j}}^{g}(t)\right\} \right\rangle \left\langle g,\left\{ \psi_{n_{j}}^{g}(t)\right\} \right|
\end{eqnarray}
where
\begin{equation}
\left|e,\left\{ \psi_{n_{j}}^{e}(t)\right\} \right\rangle =\left|e\right\rangle \otimes e^{-iH^{e}t}\left|\left\{ \psi_{n_{j}}^{e}\right\} \right\rangle 
\end{equation}
and similarly for the ground state.

Again, to find the average energy over time for arbitrary system distributions,
we consider the evolution of the operators 
\begin{eqnarray}
a_{e,j}(t) & = & e^{iH^{e}t}a_{j}e^{-iH^{e}t},\label{eq:a_e_j}\\
a_{g,j}(t) & = & e^{iH^{g}t}a_{j}e^{-iH^{g}t}.\label{eq:a_g_j}
\end{eqnarray}
and their adjoints under the Heisenberg picture. Since the multiple
modes of the oscillator are orthogonal, it's obvious we can generalize
Eqs.~(\ref{eq:Hei_a_e})-(\ref{eq:Hei_a_g}) to
\begin{eqnarray}
\dot{a}_{e,j}(t) & = & -i\omega a_{e,j}(t)-i\eta_{j},\\
\dot{a}_{g,j}(t) & = & -i\omega a_{g,j}(t)+i\eta_{j}.
\end{eqnarray}
The evolutions according to these equations will give rise to the
average energy
\begin{eqnarray}
\left\langle H_{S}+H_{I}\right\rangle _{C} & = & \biggl[\Omega+\sum_{j}2\eta_{j}\biggl(\alpha_{j}\cos(\omega_{j}t-\phi_{j})\nonumber \\
 &  & +\frac{\eta_{j}}{\omega_{j}}(P_{e}-P_{g})(\cos\omega_{j}t-1)\biggl)\biggl]\sigma_{z}\label{eq:avg_multi}
\end{eqnarray}
over the orthogonal controller basis $\left\{ \left|\psi_{n_{j}}^{e}\right\rangle ,\left|\psi_{n_{j}}^{g}\right\rangle \right\} $.
Similarly, $\alpha_{j}$ and $\phi_{j}$ are the initial amplitude
and phase of the $j$-th mode.

Eq.~(\ref{eq:avg_multi}) looks like a simple extension to Eq.~(\ref{eq:avg_single})
in its form. However, by reshuffling the terms, we can arrange it
to become a Fourier series about time $t$ with nonzero coefficients
in both the sines and the cosines
\begin{eqnarray}
\left\langle H_{S}+H_{I}\right\rangle _{C} & = & \sigma_{z}\biggl\{\Omega-(P_{e}-P_{g})\sum_{j}\frac{2\eta_{j}^{2}}{\omega_{j}}+\sum_{j}2\eta_{j}\nonumber \\
 &  & \biggl[\biggl(\alpha_{j}\cos\phi_{j}+\frac{\eta_{j}}{\omega_{j}}(P_{e}-P_{g})\biggl)\cos\omega_{j}t\nonumber \\
 &  & +\alpha_{j}\sin\phi_{j}\sin\omega_{j}t\biggl]\biggr\}.
\end{eqnarray}
When setting this Fourier series with different coefficients, we can
obtain different cyclic waves. In other words, the multi-mode coupling
between the oscillator as the controller and the two-level system
has given us an edge of control over the average energy stored in
the system by setting different initial states for the controller. 

The typical case is when setting the Fourier series as a square wave,
for which the on-off switching of the stored energy occurs. Consider
that we let
\begin{equation}
\left\{ \begin{array}{l}
2\eta_{j}\alpha_{j}\sin\phi_{j}=\frac{A}{\omega_{j}}\\
2\eta_{j}\left(\alpha_{j}\cos\phi_{j}+(P_{e}-P_{g})\frac{\eta_{j}}{\omega_{j}}\right)=0
\end{array}\right.
\end{equation}
to furnish a sinc function for the sine series and zero out the cosine
series. Then the square wave of height $A$ would be realized if $\omega_{j}=(2j-1)\omega_{0}$
are the odd harmonics of some fundamental frequency $\omega_{0}$
and
\begin{eqnarray}
\alpha_{j} & = & \frac{\sqrt{4(P_{e}-P_{g})^{2}\eta_{j}^{4}+A^{2}}}{2\eta_{j}(2j-1)\omega_{0}},\\
\phi_{j} & = & -\tan^{-1}\left(\frac{A}{2(P_{e}-P_{g})\eta_{j}^{2}}\right).
\end{eqnarray}
We can observe we conclude that the amplitudes $\alpha_{j}$ determines
the frequency of the Fourier series while the phases $\phi_{j}$ determines
the height of the square wave or the amount of energy being transfered
in and out of the system.

Finally, following the nomenclature of quantum thermodynamics, we
can define a spectral density function
\begin{equation}
J(\omega)=\sum_{j}\frac{4(P_{e}-P_{g})^{2}\eta_{j}^{4}+A^{2}}{4\eta_{j}^{2}\omega_{j}^{2}}\delta(\omega-\omega_{j})
\end{equation}
for the oscillator controller as a heat reservoir. This system-specific
reservoir will supply energy to the system such that it will undergo
energy cycling with the system at period $T=2\pi/\omega_{0}$, during
which half of the time the system will be turned on to attain a higher
energy and half of the time the system will be turned off to a lower
energy state. The distribution of the spectral density depends on
the coupling strengths $\eta_{j}$ as well as the desired population
distribution $\{P_{e},P_{g}\}$ of the two-level systems.

\begin{figure}
\includegraphics[bb=1bp 182bp 600bp 600bp,clip,width=10cm]{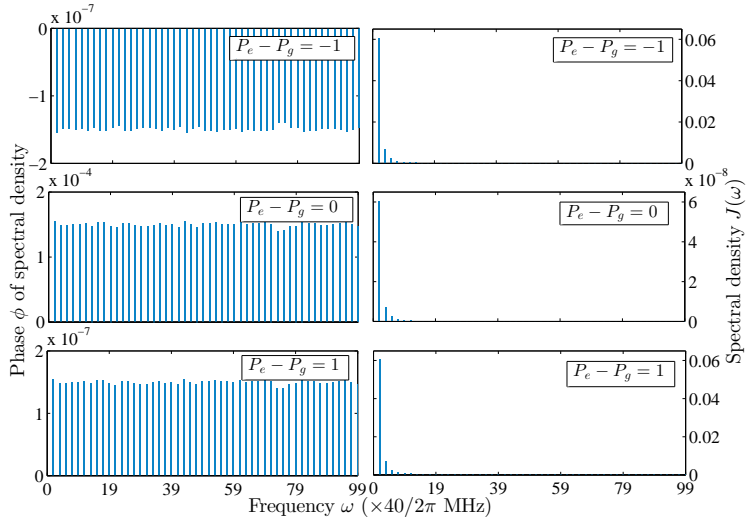}\protect\caption{Spectral density distributions $J(\omega)$ for the first 50 odd harmonics
of base frequency $\omega_{0}$ with population inversion $P_{e}-P_{g}$
at no inversion, half inversion, and full inversion, respectively.\label{fig:spec_density}}
\end{figure}

In Fig.~\ref{fig:spec_density}, the spectral density is plotted
for three values of $P_{e}-P_{g}$: -1 for no population inversion,
0.001 for half inversion (uniform Bernoulli distribution), and 1 for
full population inversion. The targeted amplitude $A$ is $30/2\pi$
MHz. The base frequency $\omega_{0}$ is set the experimentally accessible
detuning of $40/2\pi$ MHz between the cavity and a driving laser.
The first fifty odd harmonics over this base frequency are considered
for the heat reservoir. We assume the spectral density of the coupling
strength adopt a normal distribution with variance $100/2\pi$ kHz
about the mean $10/2\pi$ MHz. We can notice that to realize the on-off
switching effect, the oscillator is generally not very difficult to
prepare. No matter what the amount of inversion the two-level system
retains, the associated spectral density of the oscillator as a heat
reservoir has fairly uniform phase distributions. As for the magnitude
$J(\omega)$, we observe it is sufficient to produce only the first
few harmonics for the purpose of heat transfer. In particular, for
the half-inversion case, the amplitude of $J(\omega)$ is almost negligible
across all harmonics since the uniform Bernoulli distribution at the
two-level system naturally induces the energy exchange with the oscillator
controller.

\section{Thermodynamic Cycle of cavity optomechanical system\label{sec:Thermodynamic-Cycle}}

With the on-off effect shown on an oscillator-coupling two-level system,
we now turn eventually to the study of a movable mirror in an optomechanical
cavity associated with the thermodynamic energy transfer. The previously
studied two-level system $\sigma_{z}$ is replaced here by a pair
of annhilation and creation operators $\{b,b^{\dagger}\}$ to represent
an optical field traversed in the cavity. The effective dynamics of
the movable mirror mounted on a cantilever is determined by the flexibility
modulus of the cantilever materials. It is modeled by a multi-mode
oscillator with Hamiltonian $\sum_{j}\omega_{j}a_{j}^{\dagger}a_{j}$
and can be regarded as a heat bath under the frameworks of Feynman-Vernon~\cite{feynman63}
and Caldeira-Leggett~\cite{caldeira81}.

The motion of the mirror deforms the cavity volume, which results
in a radiation pressure proportional to the cavity photon number $b^{\dagger}b$
and the mirror displacement $x$ being fedback to the mirror~\cite{meystre85}.
Expressing the displacement $x$ in terms of the canonical conjugate
variables $a_{j}$ and $a_{j}^{\dagger}$, the total Hamiltonian reads
\begin{eqnarray}
H & = & H_{S}+H_{C}+H_{I}\nonumber \\
 & = & \Omega b^{\dagger}b+\sum_{j}\omega_{j}a_{j}^{\dagger}a_{j}+b^{\dagger}b\sum_{j}\eta_{j}(a_{j}+a_{j}^{\dagger}).
\end{eqnarray}

To examine the heat exchange process over time, we consider the density
matrix
\begin{equation}
\rho(0)=\rho_{S}(0)\otimes\rho_{C}(0)=\sum_{m}P_{m}\left|m,\left\{ \psi_{n_{j}}^{m}\right\} \right\rangle \left\langle m,\left\{ \psi_{n_{j}}^{m}\right\} \right|.
\end{equation}
to represent an initial mixed state at thermal equilibrium, where
$\left|m\right\rangle $ is the Fock eigenstate for the photon number
in the cavity and $\left|\psi_{n_{j}}^{m}\right\rangle $ is the associated
eigenstate of the mirror controller. At time $t$, the effective Hamiltonian
that drives the evolution of $\left|\psi_{n_{j}}^{m}\right\rangle $
is 
\begin{equation}
H^{m}=m\Omega+\sum_{j}\omega_{j}a_{j}^{\dagger}a_{j}+m\sum_{j}\eta_{j}(a_{j}+a_{j}^{\dagger}),
\end{equation}
which gives rise to the reduced density matrix of the mirror controller
as
\begin{equation}
\rho_{C}(t)=\sum_{m}P_{m}e^{-iH^{m}t}\left|\left\{ \psi_{n_{j}}^{m}\right\} \right\rangle \left\langle \left\{ \psi_{n_{j}}^{m}\right\} \right|e^{iH^{m}t}.
\end{equation}
Letting $\left\langle H_{S}(t)+H_{I}(t)\right\rangle _{C}=\Omega(t)b^{\dagger}b$,
we find the effective eigenenergy of the cavity system to be
\begin{equation}
\Omega(t)=\Omega+\sum_{m}P_{m}\sum_{j}\eta_{j}\left\langle \left\{ \psi_{n_{j}}^{m}\right\} \right|\left[a_{m,j}(t)+a_{m,j}^{\dagger}(t)\right]\left|\left\{ \psi_{n_{j}}^{m}\right\} \right\rangle ,
\end{equation}
where $a_{m,j}(t)$ and $a_{m,j}^{\dagger}(t)$ are the evoluted operators
in the Heisenberg picture similar to those defined in Eqs.~(\ref{eq:a_e_j})-(\ref{eq:a_g_j}),
except that the system states are here extended to all $m$ levels.

Following the same routine of the last section but assuming a continuous
spectrum for the mirror as a heat reservoir, we would arrive at
\begin{equation}
\Omega(t)=\Omega_{0}-2\int\mathrm{d}\omega\frac{\eta}{\omega_{0}}\left\{ \left(\alpha\cos\phi+\frac{\mu\eta}{\omega}\right)\cos\omega t+\alpha\sin\phi\sin\omega t\right\} ,
\end{equation}
where $\Omega_{0}=\Omega-2\mu\int(\eta^{2}/\omega)\mathrm{d}\omega$
is the renormalized energy offset when the cavity interacts with the
heat reservoir. In the equation, $\mu=\sum_{m}P_{m}m$ is the weighted
mean of the photon population across all levels in the cavity system.
To furnish the on-off effect studied in the last section, the phase
and magnitude of the spectral density should, therefore, be
\begin{eqnarray}
\phi(\omega) & = & -\tan^{-1}\frac{A\omega_{0}}{2\mu\eta^{2}},\label{eq:phi_omega}\\
J(\omega) & = & \frac{4\mu^{2}\eta^{4}+A^{2}\omega_{0}^{2}}{4\eta^{2}\omega^{2}}.\label{eq:J_omega}
\end{eqnarray}

\begin{figure}
\includegraphics[bb=105bp 250bp 562bp 535bp,clip,width=10cm]{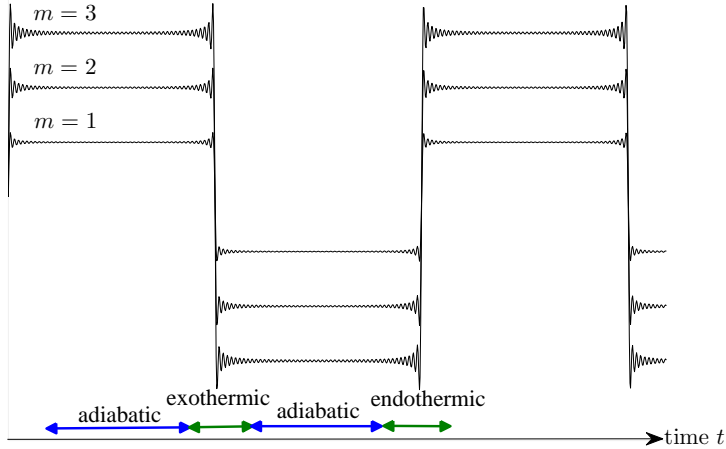}

\protect\caption{(Color online) Level diagrams of the first three Fock states of the
cavity system over time, exhibiting a cyclic energy transfer between
the system and the reservoir. Within each cycle, four thermodynamic
stroke processes can be identified.~\label{fig:cavity_sys_cycle}}
\end{figure}

If the spectral density is set so, $\Omega(t)$ will cycle about $\Omega_{0}$
with amplitude $A$ between two constant values, as illustrated in
Fig.~\ref{fig:cavity_sys_cycle}, where the reservoir is again assumed
to consist of 50 odd harmonics over a base frequency. The durations
within which $\Omega(t)$ stays constant can be identified with adiabatic
processes. During these processes, though the cavity remains interacting
with the reservoir, no energy is transferred into or out of the cavity
and all levels $m$ remain constantly spaced. Between these processes,
the cavity system either absorbs energy from the reservoir, raising
up all levels simultaneously, or infuses energy back to the reservoir,
letting itself fall back to the original levels. The absorption of
energy can be identified, thermodynamically, with an endothermic process
whereas the depletion of energy can be identified with an exothermic
process.

\section{Conclusion\label{sec:Conclusion}}

We have shown a specific thermodynamic cycle on a cavity optomechanical
system. By first demonstrating the cyclic eigenenergy of a two-level
system interacting with a single-mode oscillator, we then proves that
within the cavity system, the cavity field can act as the thermal
system relative to the multi-mode oscillating mirror acting as the
heat reservoir that controls the energy flow. By matching the spectral
density of the mirror with that of a square wave in the cyclic eigenenergy,
an on-off cycle of energy exchange can be constructed, during which
four thermodynamic stroke processes can be identified.

Therefore, a quantum mechanical system with appropriate spectral densities
can serve as a quantum thermodynamic machine. Future investigations
will focus on how the constructed thermalization processes fit within
the general framework of quantum heat engines.

\ack{}{}

The author thanks the support by FDCT of Macau under grant 013/2013/A1
and by University of Macau under grant MRG022/IH/2013/FST.

\section*{References}{}

\end{document}